# Picosecond electric-field-induced threshold switching in phase-change materials

(Peter Zalden[+,1,2], Michael J. Shu[1,3])*, Frank Chen[1,4], Xiaoxi Wu[1], Yi Zhu[5], Haidan Wen[5], Scott Johnston[3], Zhi-Xun Shen[3], Patrick Landreman[8], Mark Brongersma[8], Scott W. Fong[4], H.-S.Philip Wong[4], Meng-Ju Sher[8], Peter Jost[6], Matthias Kaes[6], Martin Salinga[6], Alexander von Hoegen[6], Matthias Wuttig[6,7] and Aaron Lindenberg[+,1,2,8]

\* These authors contributed equally
+ Corresponding authors: Peter Zalden, peter@zalden.de; Aaron Lindenberg, aaronl@stanford.edu

1) Stanford Institute for Materials and Energy Sciences, SLAC National Accelerator Laboratory, Menlo Park, CA 94025, USA
2) PULSE Institute, SLAC National Accelerator Laboratory, Menlo Park, CA 94025, USA
3) Department of Applied Physics, Stanford University, Stanford, CA 94305, USA
4) Department of Electrical Engineering, Stanford University, Stanford, CA 94305, USA
5) Advanced Photon Source, Argonne National Laboratory, Argonne, IL 60439, USA
6) I. Physikalisches Institut (IA), RWTH Aachen University, 52056 Aachen, Germany
7) JARA - Fundamentals of Information Technology, RWTH Aachen University, 52056 Aachen, Germany
8) Department of Materials Science and Engineering, Stanford University, Stanford CA 94305, USA

Many chalcogenide glasses undergo a breakdown in electronic resistance above a critical field strength. Known as threshold switching, this mechanism enables field-induced crystallization in emerging phase-change memory. Purely electronic as well as crystal nucleation assisted models have been employed to explain the electronic breakdown. Here, picosecond electric pulses are used to excite amorphous $Ag_4In_3Sb_{67}Te_{26}$. Field-dependent reversible changes in conductivity and pulse-driven crystallization are observed. The present results show that threshold switching can take place within the electric pulse on sub-picosecond time-scales - faster than crystals can nucleate. This supports purely electronic models of threshold switching and reveals potential applications as an ultrafast electronic switch.

Threshold switching is an essential mechanism in the application of phase-change materials (PCM), where electric pulses are used to thermally cycle an active material between its amorphous and crystalline states[1–3]. The amorphous phase of most chalcogenide glasses employed in this emerging technology has a conductivity that is too low to enable Joule heating to the crystallization temperature using voltages that are readily available in an electronic device[4]. However, the so-called "threshold switching" (TS) induces a breakdown in electronic resistance – for instance by three orders of magnitude[5] in the common PCM $Ge_2Sb_2Te_5$. This allows for significant heating leading to crystallization if the electric current is maintained for a sufficient duration. On the other hand, if the electric field is removed quickly after TS, the material returns to its amorphous, highly resistant state[6]. These observations have led to the development of a collection of different models to explain this effect[3,7,8]. In particular the influence of atomic rearrangements has been discussed ever since the first experimental finding of TS, with early attempts of explaining it based on field-driven nucleation effects[9]. It was also shown that TS can be explained based on entirely electronic models, coupled to the atomic temperature[5][8]. The underlying electronic transport mechanism in amorphous PCMs requires several models to explain its field dependence. At lowest field strength, Ohmic behavior is observed, whereas upon increasing field strength, the current begins to depend exponentially on the field with a sequence of Poole, Poole-Frenkel and Thermally Assisted Tunneling transport. At even higher field strength, direct tunneling of carriers becomes possible[10].

Early studies of TS in memory cells revealed the existence of a critical threshold field and a delay time between the application of an electric field and the breakdown in resistivity[11], [6]. It was found that this delay time decreases exponentially with increasing voltage – for delay times in the sub-µs regime. More recent data extend the measurement range to higher voltages[8,12] and indicate that the delay may converge toward a lower limit on the order of few nanoseconds. This would imply an upper limit in the electronically driven operation speed of phase-change memory devices. So far the shortest pulses reported for electrical excitation of a phase-change memory device based on $Ge_2Sb_2Te_5$ (GST) are of 0.5 ns duration[13]. Here we use few picosecond single-cycle electromagnetic waves with 0.5 THz center frequency to optically apply an electric field to the PCM, as shown in Fig. 1a. The impact of these pulses in a subthreshold regime, where no irreversible change occurs, has been described in earlier work[14]. To enhance the electric field strength we use electrode patterns with sub-wavelength, 1 µm gaps (see schematic drawing in Fig. 1b)[15]. Samples consist of 100 nm-thick gold bars in an interdigitated pattern, with 1 µm gap width and 10 µm electrode width, thermally evaporated and lithographically patterned on a quartz substrate. A 390 nm thick film of $Ag_4In_3Sb_{67}Te_{26}$ (AIST) is deposited by DC magnetron sputter deposition on top of this structure, filling the gaps between the gold bars. The film is capped with 20 nm of alumina grown using atomic layer deposition (ALD). The main advantage of AIST over GST for this study is its lower threshold field strength of 190 kV/cm as compared to 560 kV/cm for GST[16], while AIST is equally known to crystallize rapidly[17][18].

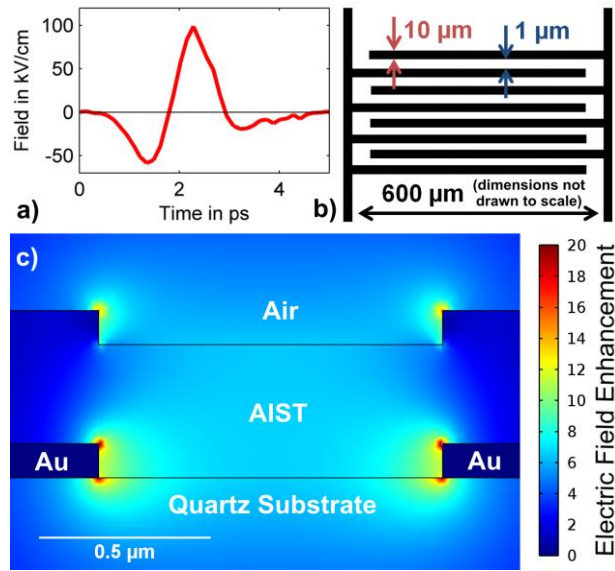

*Fig. 1: THz pulses with peak field strength of up to 100 kV/cm and picosecond duration (a) interact with an array of electrodes (b), whose gaps are filled with the PCM AIST. The electrodes enhance the electric field (c) over that observed in a bare film (capping layer not included).*

Incoming THz pulses interact with the linear electrodes enhancing the electric field inside the PCM as compared to a bare film[19]. For the given geometry with gaps of 1 µm the field is enhanced by a factor of 6.8 as compared to the field strength in AIST in the absence of electrodes - as verified experimentally and confirmed by finite element simulations presented in the methods section of the supplementary material[20]. With these gaps our experimental setup allows applying THz pulses with peak field strength of up to 480±90 kV/cm to the PCM, based on the incident field 100 kV/cm and the field enhancement accounting for reflection losses. Each THz pulse has a duration of only few picoseconds - orders of magnitude shorter than the delay times reported in literature. Each pulse drives a current that Joule heats the PCM by the temperature $\Delta T$,

$$\Delta T = \int \sigma(E(t)) \cdot E(t)^2 / c_p \, dt \qquad (1)$$

for THz conductivity $\sigma$, specific heat $c_p$, and enhanced THz electric field $E(t)$ depicted in Fig. 1a for the incoming pulse. Note that this is equivalent to absorption of THz photons by the PCM with imaginary component of dielectric function $\varepsilon'' = \sigma \cdot (\varepsilon_0 \omega)^{-1}$. With $\sigma = 2.5$ S/cm for amorphous AIST at THz frequencies[21], one concludes that every excitation pulse of 480 kV/cm heats the AIST by 0.6 K. At room temperature, the heating due to a single pulse is far below what is required to reach the crystallization temperature of 440 K[22].

We expose the device to THz pulses of 480 kV/cm peak field strength at a repetition rate of 30 Hz for 30 seconds. Even though the temperature of AIST is not expected to rise by more than 0.6 K during repetitive excitation, we observe crystallized filaments that were formed during the exposure. Crystallized material is observed by optical white light microscopy after THz exposure via the formation of highly reflective spots that did not exist in the unexposed sample (Fig. 2a)[23]. These spots also possess a higher electronic conductivity, as demonstrated by microwave impedance microscopy (MIM) data shown in Fig. 2b. The higher conductivity of the crystallized material produces a decrease in the depicted in-phase component of the reflected microwave signal. Finally, we present x-ray diffraction data with micrometer spatial resolution (Fig. 2c&d), which evidence the occurrence of Bragg diffraction characteristic of crystalline AIST at specific µm-sized positions within the gap. Fig. 2c maps the integrated diffracted intensity from gold around $q = 2.67$ Å$^{-1}$ [24] and reveals the end of one of the gold electrodes. Mapping the same area based on the integrated diffraction from crystalline AIST around $q = 2.02$ Å$^{-1}$ [17,22] results in Fig. 2d, which demonstrates a µm-sized area of polycrystalline AIST at the corner of the electrode. This is not a single grain, as evidenced by the occurrence of multiple diffraction spots with the same $q$ (not shown). In this image, the crystalline filament appears elongated along the slit due to the elliptical x-ray footprint on the sample of 5 µm long diameter (350 nm vertical diameter). These property changes in optical reflectivity, electronic conductivity and lattice ordering evidence crystallization after repetitive THz excitation. From the fact that crystallization occurs, we can derive that crystallization temperatures were reached. This implies, following eq. 1, that the electronic conductivity must increase by more than two orders of magnitude during the application of the THz pulse. Our results therefore show that TS takes place on the picosecond timescale of the THz pulse. The field strength used in this study is comparable to that used for switching electronic devices and the electronic transport should therefore follow the same fundamental principles.

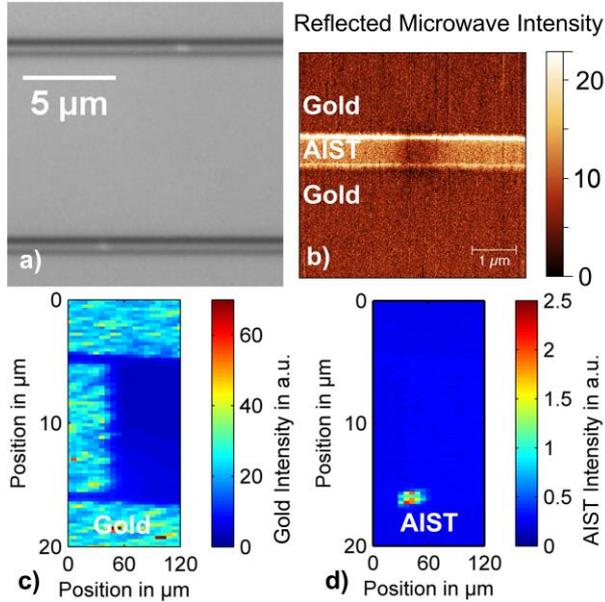

*Fig. 2: Evidence for filamentary crystallization of AIST after repetitive THz exposure (ex-situ): a) Mapping of optical white light reflectivity reveals brighter marks within the gap, b) Microwave dissipation (in-phase) signal shows a local decrease in resistance within the slit, c)+d) X-ray diffraction intensity in specific q-ranges for gold c) and AIST d), respectively, reveals local lattice ordering.*

In order to investigate the THz-induced crystallization mechanism in more detail, we have performed a THz pump/near-infrared (NIR) probe measurement of the magnitude of the THz-induced Joule heating. This measurement was carried out on similar samples, with split-ring resonators (SRR) in place of the linear electrodes. These resonator structures are depicted schematically in the inset of Fig. 3b). They confine the field to act on a well-defined area of AIST within the slit that can be probed using a microscopically focused near-infrared (NIR) pulsed beam. Fig. 3a) shows the result of a time delay scan upon repetitive THz excitation in a pump probe scheme. The electric field modulates the NIR transmission via an instantaneous electroabsorption effect while driving a current that Joule heats the material. After the excitation at about 2 ps delay, a fast relaxation mechanism is observed with time constant of 3±1 ps - most likely due to the relaxation of emitted carriers. Following the relaxation, the transmission change remains constant over at least tens of picoseconds until the heat diffuses out of the PCM. We scale this transmission change to a calibrated temperature dependent transmittance (see supplementary Fig. S4[20]) such that the NIR transmission 8 ps after THz excitation is a measure of the temperature jump in the AIST. The resulting temperature jump is shown in Fig. 3b), measured repetitively while cycling the THz pulse fluence to a level just below the onset of crystallization.

These data reveal that THz pulses increase the temperature by tens of Kelvin. Note that the experiment determines an average temperature over the 8 µm diameter of the probe beam, whereas Fig. 2 shows that only few µm-sized conducting channels are formed inside the PCM. Joule heating is significantly higher in these filaments than in the surrounding material, and therefore the experimentally obtained temperature underestimates the peak temperature of the filament. Considering this, we can estimate the temperature of the switched filament by scaling the experimental temperature jump by the ratio of the probe beam diameter to the conducting channel size. Based on the >15 K temperature jumps observed and the formation of 1 µm wide filaments along the 8 µm beam diameter, this gives temperature jumps of ~120 K comparable to the crystallization temperature in this just-below-crystallization regime. Therefore, our experimental data are consistent with a Joule heating-based thermal model of nucleation and growth to explain crystallization. It is worth mentioning that the active volume in our devices is well above the lower limit of percolation theory[25].

The data in Fig. 3b) also reveal that upon exceeding a specific threshold field, whose absolute value will be discussed later, the temperature jump starts to increase more dramatically, with a strongly non-linear dependence on the applied electric field. Since the THz fluence is proportional to the square of the peak field, it is evident from eq. 1 that this factor of ten increase in slope (heating/fluence) implies that the conductivity of the probed material increases by at least an order of magnitude to >25 S/cm when approaching crystallization conditions. Commonly, threshold switching is accompanied by at least three orders of magnitude increase in conductivity over the DC value of 0.025 S/cm. In the present case the increase appears smaller for two reasons. Firstly, an additional phonon absorption mechanism of 2.5 S/cm conductivity is present at THz frequencies[21], acting as a field-independent background. The effect of this constant phonon contribution is depicted as a red line in Fig. 3b), obtained from fitting the constant conductivity data at lowest THz fluence. Further experimental evidence for this low-field THz conductivity/absorption mechanism of amorphous AIST can be found in part D of the supplementary information[20]. Secondly, the probe beam averages over inhomogeneously switched material, reducing the derived modulation in the conductivity of the switched filament and implying conductivities of order 250 S/cm inside the filament. Therefore, our data provide evidence for a THz-induced increase of the electronic conductivity by more than three orders of magnitude over the DC

value - just as commonly observed during TS. This increase in electronic conductivity is larger than the field dependence of the sub-threshold conductivity in AIST[10] and therefore confirms that an electronic instability occurs during the picosecond THz excitation.

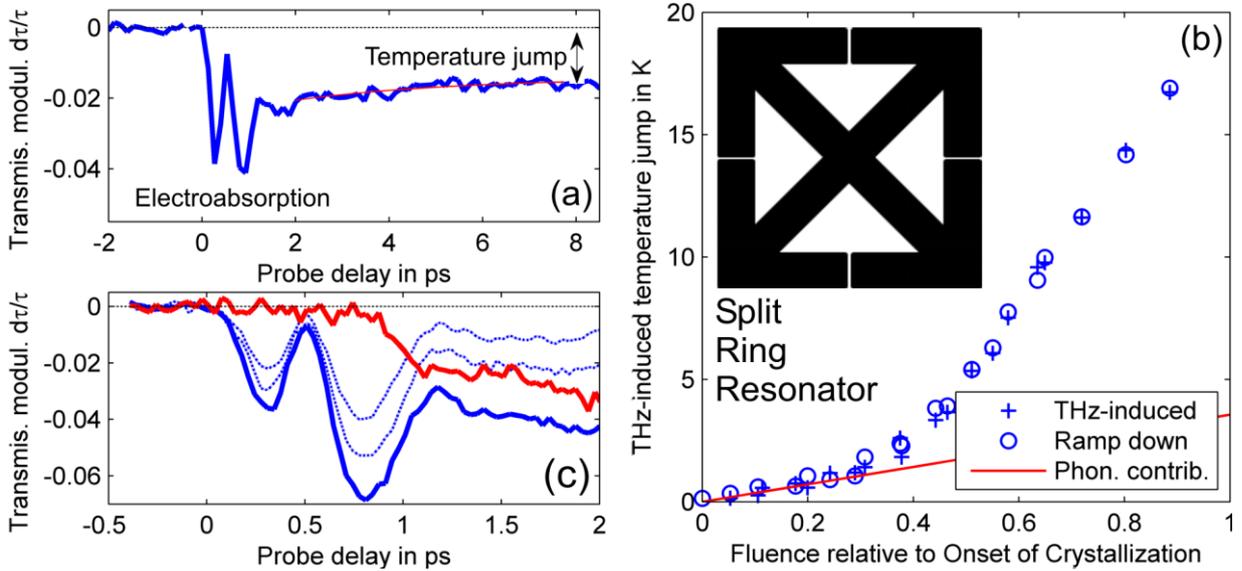

*Fig. 3:* (a) THz pulses are absorbed in the PCM and induce temperature jumps that modulate its optical transmission after carrier relaxation (blue curve with red refined exponential decay). This long-lived transmission change is due to Joule heating of the PCM. (b) At low THz fluences this temperature jump also scales linearly with the square of the electric field, dominated by a phonon-absorption mechanism. At higher peak fields, however, a one order of magnitude increase in slope (conductivity) marks the onset of TS. (c) The instantaneous field-induced modulation due to electroabsorption scales linearly with the square of the electric field (intensity) and can be subtracted from the data at highest fluence (thick curve), revealing the ultrafast onset of the non-linear Joule heating mechanism (red curve).

Another characteristic feature of TS is the reversibility below the onset of crystallization. Fig. 3b) also provides evidence for this reversibility, because the data points were obtained in a pump-probe experiment and therefore TS takes place during single pulse excitation. It is worth mentioning that the data in Fig. 3b) cannot be explained by gradual crystallization of the PCM, because the initial low-conductivity regime is retained once the peak field is decreased again. Furthermore, at stronger fields when the device crystallizes, the formation of a permanent crystalline conduction path suppresses the field enhancement of the SRR, which is seen experimentally as an immediate weakening of the pump-probe effect (not shown).

The timescale of the THz-induced transmission change (Fig. 3a) is hidden underneath the electroabsorption signal, which scales with the square of the electric field[14][26] and dominates at weak excitation conditions. Fig. 3c) shows delay scans at three different THz fluences (blue curves) of 0.4, 0.55 and 0.7 times the fluence at the onset of crystallization. Based on these data it becomes possible to subtract the electroabsorption effect, revealing the ultrafast heating mechanism in the remaining signal (red curve), previously hidden in the raw data (thick blue curve). This scan reveals the electronic breakdown mechanism associated with TS on a sub-picosecond timescale.

It is well known that the electric field required for TS increases for short pulse durations[11], [26]. With the SRR structures, the confinement of the electric field precludes an experimental determination. Therefore, we performed the same pump-probe experiment on the linear electrode structures discussed earlier, where the absolute field strength can be simulated and measured for a given THz fluence based on the sub-threshold homogeneous heating of the phase-change material. We find that the absolute threshold field for AIST is 230±60 kV/cm - not significantly deviating from literature data using DC pulses (190 kV/cm). Details of this determination can be found in part B of the supplementary information[20]. It is worth mentioning that the literature value was obtained with unipolar pulses, whereas for bipolar THz pulses the 1.5 times higher peak-to-peak field swing may be relevant for inducing TS. Presently, no comparative study of AC vs. DC threshold field strengths is known to the authors. At peak field strength of 340 kV/cm and above, crystallization is found to take place in the linear electrodes. This ratio of threshold fluence to onset of crystallization $(230/340)^2 = 0.46$ is very similar to the ratio found for the SRR devices (see Fig. 3b).

A summary of models to describe TS is beyond the scope of this manuscript and therefore we will compare our results to a very recent and comprehensive model[10]. It is based on a combination of linear and non-linear electronic transport mechanisms, which are dominating depending on the applied field strength. Extrapolating this quantitative model for AIST including direct tunneling to ambient temperature of 300 K, one obtains a total density of emitted carriers on the order $5*10^{20}$ cm$^{-3}$ for a THz waveform E(t) of 360 kV/cm peak field strength. This value is based on an assumed mobility of 1 cm$^2$ (Vs)$^{-1}$ and a high-field slope $F_c$ of 9.5±0.5 V/µm, whose uncertainty dominates that of the resulting carrier concentration. The absolute conductivity resulting from this estimate is on the order 100 S/cm, in good agreement with the present results.

One alternative model proposed to explain TS relies on a field-driven nucleation mechanism, where the electric field induces the transient formation of subcritical crystalline nuclei with orders of magnitude higher conductivity[9,27]. Based on our findings we can rule out models for TS that are based on nucleation processes, because we show that the sub-picosecond timescale of TS is shorter than what is required for nucleation of crystalline material: The nucleation rate is limited by the rate at which atoms vibrate and attempt to nucleate. In case of AIST, within a picosecond only less than five atomic vibrations are possible for each atom[28], which is insufficient to establish the periodic arrangement of several atoms in a crystal. Therefore, TS must be of solely electronic origin. In such a model, the lower limit for the time delay associated with TS becomes negligible as compared to the nanosecond thermal crystallization mechanism if sufficiently high field strengths are employed. Consequently, this minimum delay associated with TS is expected to be significantly shorter than a picosecond.

To support the consistency of the overall mechanism, we briefly discuss its energetics based on the sample with linear electrodes. At the onset of crystallization, each THz pulse induces a field strength of 340 kV/cm inside AIST. This value is obtained, taking into account the full dielectric functions of all materials involved. Simulations also show that this field is not diminished in the presence of narrow conducting filaments with conductivity of order $10^3$ S/cm, comparable to the conductivity of the crystalline phase (see supplementary material, Fig. S6[20]). Thus energy is effectively channeled through the electrodes toward the conductive filament. Note that this high conductivity, required for heating to the crystallization temperature, occurs at an applied field significantly above the TS voltage as seen in Fig. 3b). These filaments absorb an energy density $\sigma E^2 t = 150$ J cm$^{-3}$, enough to heat the material to the crystallization temperature, given the specific heat of AIST of 1.27 J cm$^{-3}$ K$^{-1}$ [29]. In order to heat one filament of 1 µm width to the crystallization temperature, this requires absorption of 70 pJ of energy, much less than the 45 nJ of THz energy incident on the area of each slit and half of both its neighboring electrode bars. Therefore, we conclude that a model based on purely thermal crystallization can explain the experimental results.

In conclusion, we have demonstrated that an amorphous PCM crystallizes under repetitive excitation with single-cycle THz pulses of few picosecond duration and sufficient field strength. We provide experimental evidence for most of the characteristic features of threshold switching: The formation of conducting filaments through highly non-linear conduction mechanisms, a threshold field that is in reasonable agreement with literature values and the reversibility at above threshold, but sub-crystallization conditions. In the initially amorphous device, THz pulses induce TS during the picosecond excitation. TS increases the conductivity and the resulting higher current heats the material locally above the crystallization temperature. Upon repetitive excitation, atomic rearrangements toward the crystalline phase can take place. Due to the thermodynamically irreversible nature of crystallization, these modifications are retained after the material has cooled down by heat transport into the substrate[30], [31]. The observation of picosecond TS is particularly important for ovonic switches, sometimes used as pre-selectors in memory devices[32], which then allow sub-picosecond access to the specific memory cell. In this sense, TS allows the design of ultrafast electronic switches.

Research was supported by the U.S. Department of Energy, Basic Energy Sciences, Materials Sciences and Engineering Division. MW gratefully acknowledges support by the Deutsche Forschungsgemeinschaft through SFB 917. PZ gratefully acknowledges funding from the Alexander von Humboldt Foundation. The use of the Advanced Photon Source was supported by the U.S Department of Energy, Office of Science, Office of Basic Energy Sciences, under Contract No. DE-AC02-06CH11357. MS and MK acknowledge funding from the DIASPORA project of the FP7-IAPP Marie-Curie Action by the

European Commission. SWF and HSPW are supported in part by the Stanford Non-Volatile Memory Technology Research Initiative (NMTRI). The MIM work was supported by NSF DMR 1305731.

[1] H. P. Wong, S. Raoux, S. Kim, J. Liang, J. P. Reifenberg, B. Rajendran, M. Asheghi, K. E. Goodson, *Proc. IEEE* **2010**, *98*, 2201.

[2] M. Wuttig, N. Yamada, *Nat. Mater.* **2007**, *6*, 824.

[3] D. Ielmini, Y. Zhang, *J. Appl. Phys.* **2007**, *102*, 054517.

[4] G. W. Burr, M. J. Breitwisch, M. Franceschini, D. Garetto, K. Gopalakrishnan, B. Jackson, C. Lam, A. Luis, *J. Vac. Sci. Technol. B* **2010**, *28*, 223.

[5] A. Redaelli, A. Pirovano, F. Pellizzer, a. L. Lacaita, D. Ielmini, R. Bez, *IEEE Electron Device Lett.* **2004**, *25*, 684.

[6] A. E. Owen, J. M. Robertson, *IEEE Trans. Electron Devices* **1973**, *20*, 105.

[7] M. Nardone, M. Simon, I. V. Karpov, V. G. Karpov, *J. Appl. Phys.* **2012**, *112*, 071101.

[8] M. Le Gallo, A. Athmanathan, D. Krebs, A. Sebastian, *J. Appl. Phys.* **2016**, *119*, 025704.

[9] S. R. Ovshinsky, *Phys. Rev. Lett.* **1968**, *21*, 1450.

[10] M. Kaes, M. Le Gallo, A. Sebastian, M. Salinga, D. Krebs, *J. Appl. Phys.* **2015**, *118*, 135707.

[11] S. R. Ovshinsky, *Phys. Rev. Lett.* **1968**, *21*, 1450.

[12] M. Wimmer, M. Salinga, *New J. Phys.* **2014**, *16*, 113044.

[13] D. Loke, T. H. Lee, W. J. Wang, L. P. Shi, R. Zhao, Y. C. Yeo, T. C. Chong, S. R. Elliott, *Science* **2012**, *336*, 1566.

[14] M. Shu, P. Zalden, F. Chen, B. Weems, I. Chatzakis, F. Xiong, R. Jeyasingh, M. Hoffmann, E. Pop, P. Wong, M. Wuttig, A. Lindenberg, *Appl. Phys. Lett.* **2014**, *104*, 251907.

[15] M. Liu, H. Y. Hwang, H. Tao, A. C. Strikwerda, K. Fan, G. R. Keiser, A. J. Sternbach, K. G. West, S. Kittiwatanakul, J. Lu, S. a Wolf, F. G. Omenetto, X. Zhang, K. a Nelson, R. D. Averitt, *Nature* **2012**, *487*, 345.

[16] D. Krebs, S. Raoux, C. T. Rettner, G. W. Burr, M. Salinga, M. Wuttig, *Appl. Phys. Lett.* **2009**, *95*, 2101.

[17] T. Matsunaga, J. Akola, S. Kohara, T. Honma, K. Kobayashi, E. Ikenaga, R. O. Jones, N. Yamada, M. Takata, R. Kojima, *Nat. Mater.* **2011**, *10*, 129.

[18] P. Zalden, A. von Hoegen, P. Landreman, M. Wuttig, A. M. Lindenberg, *Chem. Mater.* **2015**, *27*, 5641.

[19] M. a. Seo, H. R. Park, S. M. Koo, D. J. Park, J. H. Kang, O. K. Suwal, S. S. Choi, P. C. M. Planken, G. S. Park, N. K. Park, Q. H. Park, D. S. Kim, *Nat. Photonics* **2009**, *3*, 152.

[20] W.-P. Hsieh, P. Zalden, M. Wuttig, A. Lindenberg, W. L. Mao, *Appl. Phys. Lett.* **2013**, *103*, 191908.

[21] F. Kadlec, C. Kadlec, P. Kužel, *Solid State Commun.* **2012**, *152*, 852.

[22] W. K. Njoroge, M. Wuttig, I. Introduction, *J. Appl. Phys.* **2001**, *90*, 3816.

[23] K. Shportko, S. Kremers, M. Woda, D. Lencer, J. Robertson, M. Wuttig, *Nat. Mater.* **2008**, *7*, 653.

[24] H. E. Swanson, E. Tagte, in *Natl. Bur. Stand. Circ. 539*, **1953**, pp. 1–96.

[25] M. Simon, M. Nardone, V. G. Karpov, I. V. Karpov, *J. Appl. Phys.* **2010**, *108*, 4514.

[26] G. Weiser, U. Dersch, P. Thomas, *Philos. Mag. Part B* **1988**, *57*, 721.

[27] I. V. Karpov, M. Mitra, D. Kau, G. Spadini, Y. a. Kryukov, V. G. Karpov, *Appl. Phys. Lett.* **2008**, *92*, 173501.

[28] G. C. Sosso, S. Caravati, M. Bernasconi, *J. Phys. Condens. Matter* **2009**, *21*, 095410.

[29] P. Zalden, K. Siegert, S. Rols, H. E. Fischer, F. Schlich, T. Hu, M. Wuttig, *Chem. Mater.* **2014**, *26*, 2307.

[30] P. Zalden, A. von Hoegen, P. Landreman, M. Wuttig, A. M. Lindenberg, *Chem. Mater.* **2015**, *27*, 5641.

[31] J. Siegel, A. Schropp, J. Solis, C. N. Afonso, M. Wuttig, *Appl. Phys. Lett.* **2004**, *84*, 2250.

[32] M. Anbarasu, M. Wimmer, G. Bruns, M. Salinga, M. Wuttig, *Appl. Phys. Lett.* **2012**, *100*, 143505.

# Supplementary Material:
# Picosecond electric-field-induced threshold switching in phase-change materials

(Peter Zalden[+,1,2], Michael J. Shu[1,3])*, Frank Chen[1,4], Xiaoxi Wu[1], Yi Zhu[5], Haidan Wen[5], Scott Johnston[3], Zhi-Xun Shen[3], Patrick Landreman[8], Mark Brongersma[8], Scott W. Fong[4], H.-S.Philip Wong[4], Meng-Ju Sher[8], Peter Jost[6], Matthias Kaes[6], Martin Salinga[6], Alexander von Hoegen[6], Matthias Wuttig[6,7] and Aaron Lindenberg[+,1,2,8]

* These authors contributed equally
+ Corresponding authors: Peter Zalden, peter@zalden.de; Aaron Lindenberg, aaronl@stanford.edu

1) Stanford Institute for Materials and Energy Sciences, SLAC National Accelerator Laboratory, Menlo Park, CA 94025, USA
2) PULSE Institute, SLAC National Accelerator Laboratory, Menlo Park, CA 94025, USA
3) Department of Applied Physics, Stanford University, Stanford, California 94305, USA
4) Department of Electrical Engineering, Stanford University, Stanford, California 94305, USA
5) Advanced Photon Source, Argonne National Laboratory, Argonne, IL 60439, USA
6) I. Physikalisches Institut (IA), RWTH Aachen University, 52056 Aachen, Germany
7) JARA - Fundamentals of Information Technology, RWTH Aachen University, 52056 Aachen, Germany
8) Department of Materials Science and Engineering, Stanford University, Stanford CA 94305, USA

**Outline**

A) Methods (Static characterization tools and the Pump-Probe Setup)
B) Determination of Field Enhancement through the electrode structures
C) THz-induced crystallization
D) Conductivity spectrum of amorphous AIST

### A.1) X-ray microdiffraction

Hard x-ray microdiffraction experiments were carried out at beamline 7-ID-C of the Advanced Photon Source. Incoming x-rays of 11 keV photon energy were focussed by a Fresnel zone plate to 350 nm vertical and 1.2 µm horizontal FWHM. The sample was placed on a piezo-driven translation stage to scan diffraction from different locations along the gaps, aligned horizontally. The sample was tilted against the propagation of x-rays so that a reflection with $2\theta = 27.7°$ would be observed as symmetric reflection in the horizontal plane. An x-ray area detector (Pilatus 100K) was employed to image the diffraction pattern from each location over a q-range from 1.5 to 5.0 Å$^{-1}$. 2D diffraction intensity maps were obtained by integrating over a given q-range while scanning the sample position.

### A.2) Impedance Microscopy

Microwave impedance microscopy (MIM) probes spatial variations of the complex sample impedance. A continuous microwave signal is incident on an AFM tip and changes in the amplitude and phase of the reflected signal are detected[15],[16]. The in-phase component of the reflected signal is dominated by the local resistivity, whereas the out-of-phase component detects spatial changes in the tip-sample capacitance. In this case measurements were done at 1 GHz.

### A.3) Optical Microscopy

An optical microscope with confocal illumination was used to probe the local reflectivity of the sample. Fig. S1 depicts resonator structures before and after THz exposure.

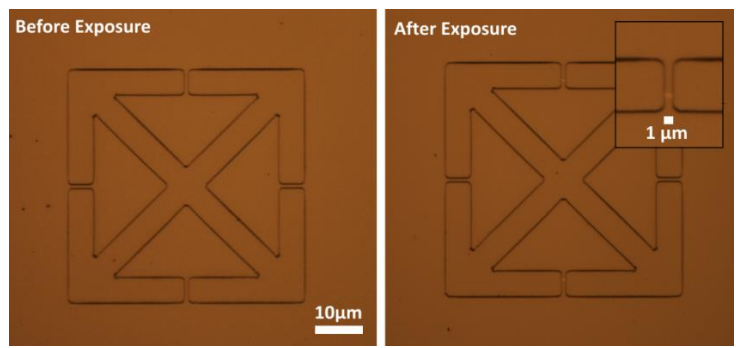

**Fig. S1:** Optical microscope image of one of the SRR structures topped with AIST. Left: Before THz exposure. Right and zoomed-in image: Filamentary crystallization is seen in the top and bottom gaps after exposure to sufficiently strong THz fields.

### A.4) The THz pump-optical probe setup

The entire experimental setup is depicted schematically in Fig. S2. A Ti:Sa regenerative amplifier system with 960 Hz repetition rate and 2 mJ pulse energy is used to (a) seed a Ti:Sa multipass amplifier that generates 10 mJ pulses of 800 nm wavelength at 30 Hz repetition rate and (b) drive an optical parametric amplifier (OPA) to generate optical pulses of 1500 nm wavelength. The output from the OPA is filtered so that spectral components are observed only at 1500 nm. Both optical pulses at 800 and 1500 nm have pulse durations of 70 fs. The 800 nm pulses are diffracted off of a grating to tilt the pulse front and generate THz radiation by group velocity matching in a single crystal of LiNbO$_3$ [1]. This technique generates single cycle electromagnetic pulses with 0.5 THz center frequency and linear polarization. They are used to excite a sample mounted at the focus of the THz optics. To probe the impact of the THz waveform, 1500 nm pulses are sent through an optical delay line and collinearly arrive at the sample location. Their transmitted intensity is determined with an InGaAs photodiode that is read out by a lock-in amplifier. In case of a fluence scan, the first of the two polarizers in the THz beampath is rotated, thereby reducing the transmitted field strength while the second polarizer keeps the output polarization unchanged.

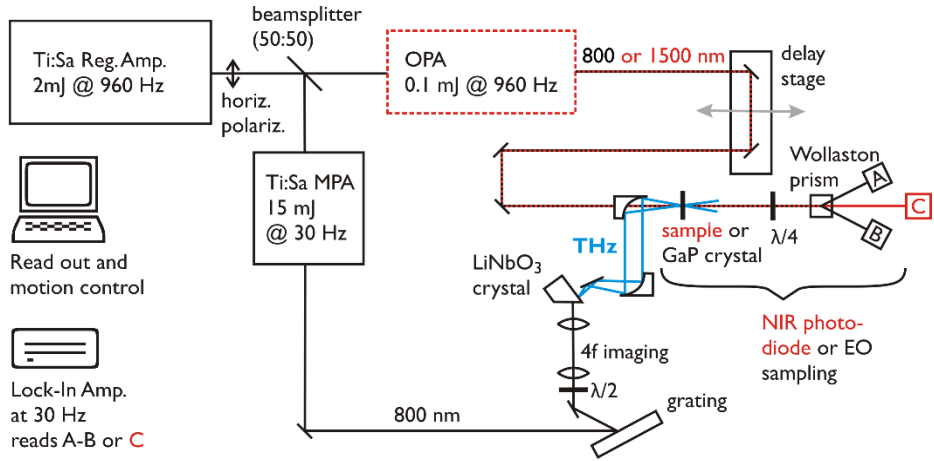

**Fig. S2:** Experimental setup for the THz excitation of phase-change materials (sample). Incoming THz field strength at the focus location can be determined by replacing the sample with a GaP crystal to perform electro-optic (EO) sampling. Temperature jumps can be measured using an OPA and determine the 1500 nm transmission of the sample (beam path shown in red).

To determine the waveform of THz pulses, the EO sampling technique is employed [2] using a 110-cut GaP crystal with 250 µm thickness. The electric field induces birefringence in the crystal, which can be detected optically. Because the optical properties of GaP are more well-known at 800 nm, we use optical pulses at 800 nm to transmit through the same part of the crystal. The birefringence is detected by sending the modulated optical pulses through a quarter waveplate and splitting them into horizontally and vertically polarized components. Each of these components is detected by a Si diode, whose differential signal $\Delta I = I_A - I_B$ is read by the lock-in amplifier. The electric field $E_{THz}$ of the waveform can be determined by recording this modulation $\Delta I$ while scanning the optical delay stage. The following equation holds for $\Delta I/I < \sim 0.2$,

$$E_{THz} = \frac{\Delta I}{I} \frac{c}{\omega n^3 r_{41} l}$$

where $I$ is the intensity of the probe beam, c is the speed of light, $\omega$ the frequency of the 800 nm probe beam, $n$ is the refractive index of GaP at 800 nm, $r_{41}$ is the electro-optic coefficient at THz frequencies and $l$ is the thickness of the GaP crystal. The uncertainty associated with the electro-optic coefficient of GaP is quite large, even though it is commonly used as 0.97 pm/V – its value at 100 THz (see [3] and [4], p. 47 with references). Other extrapolations suggest it could be almost a factor two smaller [5]. In the absence of reliable data around 1 THz, we use the common value for consistency with earlier work [3]. The present value leads to a good agreement between the THz induced temperature jump observed in the experiment and the temperature jump expected from Joule heating based on the resulting absolute field strength (see Fig. S2). Therefore, the determination of absolute field strength is subject to systematic uncertainties that we find to be below 20%. The refractive index of GaP is 3.18 [6] with ±1% uncertainty. Relative changes in the electric field strength, however, can be determined to better than

0.1%. In the present setup, the maximum modulation $\Delta I/I$ corresponds to a peak field strength of the incoming pulses of 100 kV/cm, 1.17 times higher than the field strength measured using EO sampling, taking into account the reflection losses from the surface of the GaP crystal.

### A.5) Temperature dependence of NIR transmission

The same sample coated with AIST was used to determine the optical transmission at 1500 nm while increasing the temperature (see Fig. S4). The measurement was performed on an area of the film without electrode structures. No significant differences were observed in data points taken during heating of the sample from ambient to 60°C, and upon cooling back down to room temperature. Please note that silicon as well as quartz are transparent at this wavelength, so that this curve enables transformation of transmission changes to temperature changes.

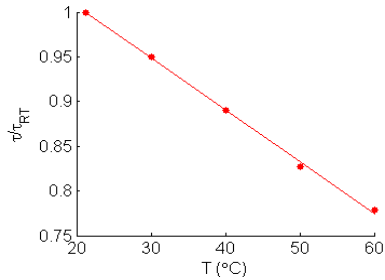

**Fig. S4:** Temperature dependence of transmission of 1500 nm beam through 390 nm AIST on quartz. The coefficient of transmission change per temperature change determined from a linear refinement is -0.0055/°C.

### B.1) Simulation and experimental verification of field enhancement

Finite element frequency-domain simulations as implemented in the RF module of COMSOL have been employed to model the propagation of the THz pulse through the electrode structure including amorphous AIST. The sample geometry as depicted in Fig. 1c of the manuscript justifies the reduction to a two dimensional model. We assume the electrodes (longer than wavelength) to be of infinite length when determining the field strength in the slits for a given incident field. In the limit of infinite slit width, the field strength in the amorphous AIST is 71% of the incoming field strength, i.e. 71 kV/cm at maximum exposure due to reflection losses. This field strength is the average over the 390 nm thick material inside the gap. Fig. S5 shows the resulting average field enhancements for different gap widths (red curve) in comparison to experimental data based on the scaling of the instantaneous electroabsorption effect, which is known to scale linearly with the peak intensity of the THz pulse [7]. Experimental data were obtained using the same electrode structures, filled with as-deposited amorphous $Ge_2Sb_2Te_5$ (GST) instead of AIST. The THz conductivity of amorphous GST (6 S/cm) is very similar [8] to that of AIST so that the results are expected to be directly comparable. The agreement between simulation and experiment further justifies the extrapolation to 1 µm gap sizes, where a field enhancement by a factor 6.8 is predicted. Deviations in the sample preparation, particularly during lithography can lead to variations in the gap size of ±0.1 µm, which in turn cause an uncertainty of the field enhancement of about 10%. The highest incident field strengths are enhanced to a maximum of 480±50 kV/cm. Note the consistency between this simulated value and a simple estimate based on the field enhancement by bare electrodes, along with the transmission for a thin film. The field $E_{film}$ inside a thin film is estimated as $E_{film} = 2E_0/(n_s + 1)$ for incident field $E_0$ (100 kV/cm) and substrate index $n_s$ (2.1 for quartz). This (65 kV/cm) is then multiplied by the field enhancement from bare electrodes, around 7, giving 450 kV/cm.

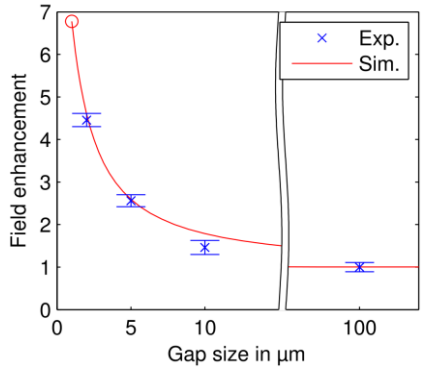

**Fig. S5:** Electric field enhancement for different slit widths in the given sample geometry. The simulated field enhancement (red curve) is based on a frequency-domain finite element simulation using COMSOL. The experimental data points were obtained for an equally thick film of amorphous $Ge_2Sb_2Te_5$. For the 1 µm slits no such effect could be measured due to damaging of the sample and therefore the field enhancement of 6.8 for this device is extracted from the simulation (red circle).

The conductivity of the material inside the gap will change upon applying electric fields of such strength. This change in conductivity could influence the field enhancement and therefore we calculated the field enhancement as a function of the conductivity of AIST inside the linear electrodes, explicitly taking into account the formation of conducting filaments. The data in Fig. S6 demonstrate that for an increasing conductivity over the initial value of 2.5 S/cm, the field enhancement first increases and then breaks down depending on the width of the filament. This implies that crystallization conditions cannot be reached by a homogeneous increase in conductivity. Only the formation of conducting filaments prevents the breakdown of the field enhancement and enables crystallization for filaments of 3 µm or less - in good agreement with the experimental results. Note that $\sigma = 10^3$ S/cm corresponds to a skin depth of $\delta = \sqrt{2/(\sigma\mu\omega)} = 1.6$ µm, comparable to the filament dimensions and consistent with a uniform absorption within the filament.

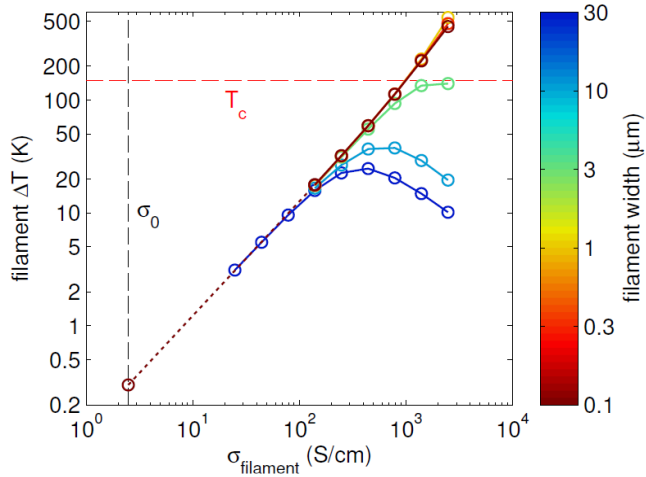

**Fig. S6:** Finite element simulation results for THz excitation of a device with a TS-induced conductive filament at 360 kV/cm peak field. The final filament temperature is computed for a given filament width and conductivity. The data point at 2.5 S/cm, where $\sigma_{filament}$ is equal to the amorphous conductivity $\sigma_0$, is equivalent to a case with no filament.

We now apply the knowledge of the absolute field strength to experimentally determine the threshold field strength of AIST upon THz excitation. Fig. S7 depicts the results of a measurement equivalent to the one of Fig. 3b) in the main text, albeit plotted as a function of the peak electric field instead of the THz fluence. In the linear geometry the location of filament formation is impossible to predict and therefore the probe beam needs to average a larger area to detect the THz induced temperature jump. In consequence, the absolute temperature jump is smaller and lateral inhomogeneities smear out the onset of TS.

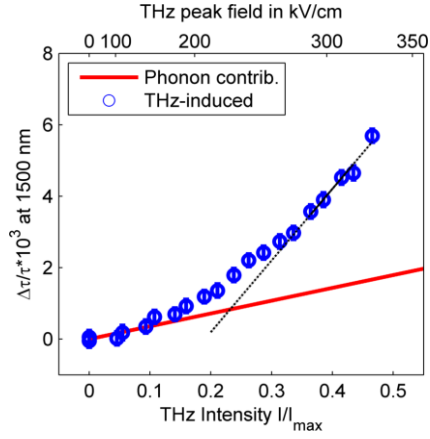

**Fig. S7:** THz-induced transmission modulation in linear electrodes as a function of the peak field strength. Crystallization sets in at 340 kV/cm, while an increase in conductivity is observed at 230±60 kV/cm (intersection of high and low-field slopes)

### C.1) THz-induced crystallization

Fig. 3 (SRRs) and Fig. S7 (linear electrodes) show the THz-induced transmission change as a function of the THz pulse fluence exciting the sample. Fig. S9a provides experimental data of the same transmission change while keeping the fluence, i.e. the peak THz field of 340 kV/cm ($I/I_{max} = 0.5$) constant. These data were obtained on an initially fresh device and show that the modulation decreases while crystallized filaments form. The amount of crystallization converges after about 600 pulses. We explain this mechanism based on the gradual formation of crystalline nuclei as the THz pulses heat localized areas beyond the crystallization temperature. The crystallized filaments then diminish the field enhancement within the electrode gaps, reducing the subsequent Joule heating (see Fig. S6). A model based on exponential decay, i.e. $d\tau/\tau = a + b \cdot \exp(-c \cdot n)$, with the number of pulses n and fit parameters a, b and c. It is important to point out that the modulation of transmission is not expected to be linearly proportional to the crystallized fraction, because crystallization of a filament short circuits neighboring electrodes and therefore reduces the overall Joule heating effect. Then, the remaining number of unshorted electrodes is reduced and the chance of further crystallization goes down – hence the empirical exponential decay to refine the experimental data.

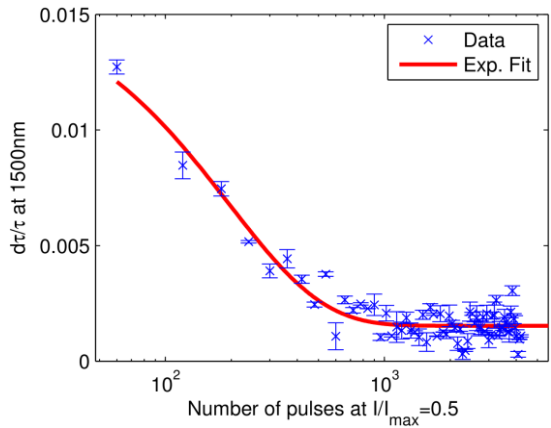

**Fig. S9a:** The THz-induced modulation of 1500 nm transmission dτ/τ, proportional to sample heating, gradually decreases as the sample crystallizes within linear electrodes.

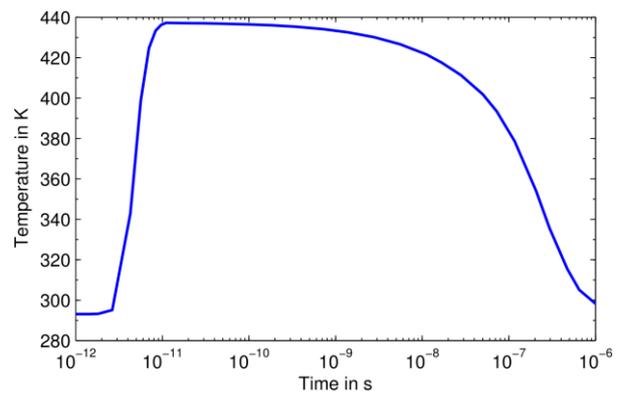

**Fig. S9b:** Transient average temperature in the linear electrode geometry. After 300 ns, the temperature jump decays to 1/e. This corresponds to an effective crystal growth time of 2 ns (in analogy with [9]).

### D.1) Temperature dependent DC and THz conductivity

Samples for DC conductivity measurements were deposited on glass substrates (Corning 1737) and measured under vacuum conditions in a custom-built instrument, which employs a Peltier stack for temperature control[17]. The conductivity was obtained in a 2-wire configuration. To ensure that the temperature dependence is not distorted by relaxation effects, all data were recorded after allowing the sample to relax for at least three hours at the highest temperature measured. The resulting data are presented in Fig. S10a), red triangles pointing down, and reveal a clear Arrhenius-like temperature dependence indicating that electronic transport in the DC regime is thermally activated with a dominating activation energy of 0.285 eV. At THz frequencies, however, the conductivity is independent of temperature within the experimental errors. These data points were extracted from temperature dependent THz spectroscopy (see C.2). This implies that the THz conductivity is not following the electronic thermal excitation of the DC conductivity. Furthermore, the THz conductivity does not show resistance drift, as expected from DC measurements after the same thermal treatment [13]. While the DC conductivity decreases by a factor two, the THz conductivity changes by less than 20%. A similar behavior was reported for $Tl_2SeAs_2Te_3$[18]. There, the electronic conductivity was reported to be frequency independent up to a plasma frequency. The orders of magnitude larger THz conductivity was found to originate from phonon absorption effects. We therefore use the same explanation for AIST and conclude that the THz absorption of AIST is dominated by phonon absorption features, giving rise to a hundred times higher total conductivity. This implies that for electronic effects to become dominant in the terahertz regime, the electronic conductivity has to increase by at least two orders of magnitude.

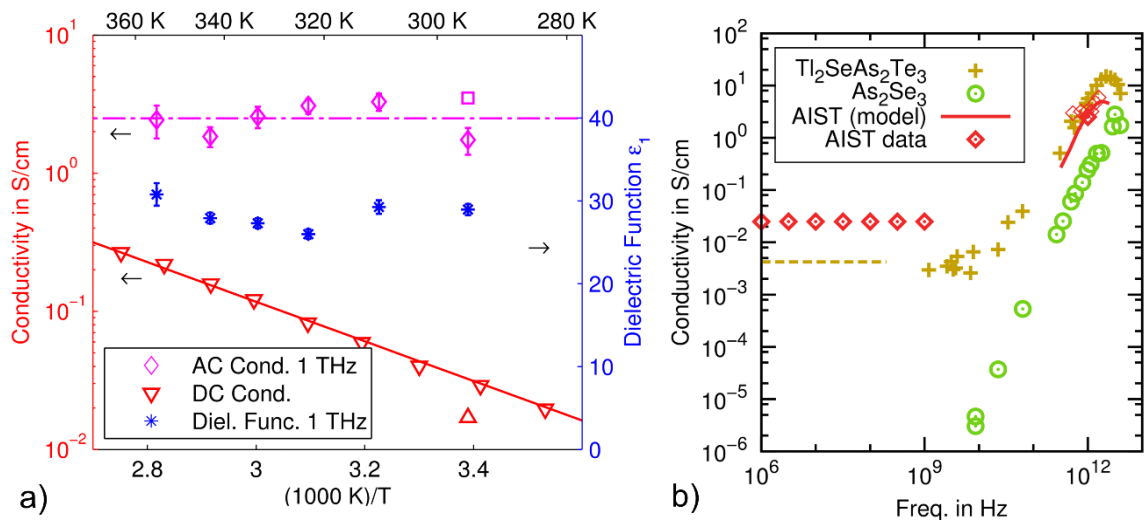

**Figure S10:** a) While the DC conductivity follows an Arrhenius behavior with 0.285 eV activation energy (red triangles, pointing down: our data; pointing up: from ref.[19]), the AC conductivity at 1 THz is temperature independent between 295 and 355 K (magenta diamonds our data, square[8]), confirming that the conductivity at 1 THz is dominated by vibrational states. b) AC conductivity of chalcogenide glasses with different DC conductivities[20]. In AIST the conductivity is flat up to at least 1 GHz, as found by impedance spectroscopy (see C.3). At higher frequencies the broad distribution of vibrational states in the glass contributes to the conductivity - an effect that is nicely resolved for the similar glass $Tl_2SeAs_2Te_3$[18].

### D.2) THz Time-Domain Spectroscopy

THz time-domain spectroscopy (TDS) was performed on the same film of AIST on quartz substrate that was used to demonstrate THz-induced crystallization. THz pulses of peak field strength around 10 kV/cm were generated using an air-plasma source [11]. They were focused on the sample at normal incidence, avoiding areas covered by electrode structures. Transmitted THz pulses were refocused and recorded using the EO sampling technique based on a 1 mm thick ZnTe crystal. The entire THz beam path was kept in a nitrogen atmosphere while heating the sample to temperatures of up to 82°C. At each temperature point, the material's dielectric constant and conductivity were determined. The parameter calculations fully account for reflections at each interface. The refractive index of the quartz substrate was taken to be 2.1 [12]. In order to take into account the effects of structural relaxation and resulting resistance drift [13], the sample was first measured under ambient conditions, annealed at 70°C and again measured at 70°C. After annealing further data points were recorded between 82°C and ambient conditions.

### D.3) Impedance Spectroscopy

The linear electrode devices were electrically connected to a vector network analyzer (VNA), measuring the imaginary and real (in-phase and quadrature) components of the reflected signal ($S_{11}$). The resulting data (Fig. S11) can be refined using a circuit model as depicted in Fig. S12. This model is based on four fit parameters: 1) an inductance of the contacting wires, $L_{wire}$ = 8 nH, 2) a frequency-independent resistance of the sample, $R_{sample}$ = 23 Ohm, 3) a capacitance for the contacts, $C_{cont}$ = 70 pF and 4) a contact resistance, $R_{cont}$ = 24 Ohm. With a film thickness of 390 nm and gap width of 2 µm in this case, the capacitance of the sample is negligible. The resulting parameters allow confirming the accuracy of the fitting in Fig. S11 by comparing with the expected resistance of the device based on its known geometry and the DC conductivity of amorphous AIST, 0.025 S/cm. The resulting resistance of 26 Ohm is in good agreement with the result from fitting (23 Ohm) and confirms the consistence of the experiment. These data confirm that the resistance of amorphous AIST can be taken as frequency independent up to at least around 1 GHz. To validate the experimental approach, the same measurement was performed on a resistor (Reference data in Fig. S11).

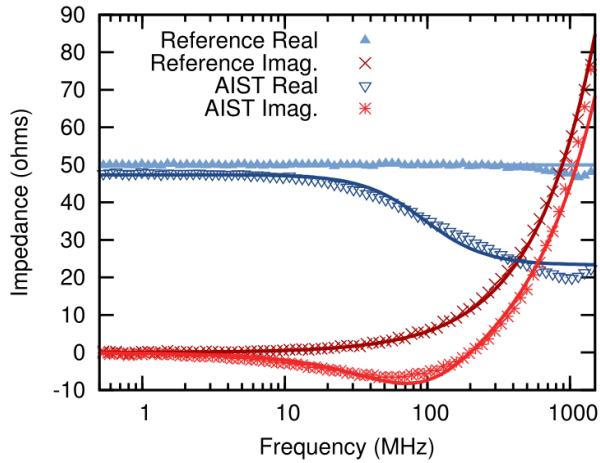

**Fig. S11:** Raw data from impedance spectroscopy with a fit based on a frequency independent resistance, both for the electrode device (AIST) as well as for a frequency independent resistor (Reference).

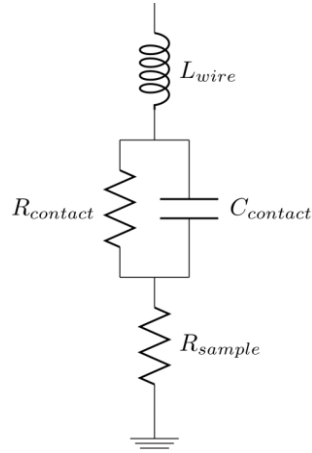

**Fig. S12:** Circuit model employed to refine the impedance spectroscopy data for our sample in planar geometry. In this geometry, the sample capacity is negligible.

### D.4) Field dependent conductivity

Fig. S13 summarizes the field dependent conductivity obtained for a phase-change line cell of given geometry. The measurement was performed by applying a DC voltage over and monitoring the current through the active element of as-deposited amorphous AIST. The applied voltage was ramped up at a rate of 3 kV (cm s)$^{-1}$. The data for AIST verify that its DC conductivity increases only by a factor of three, until threshold switching sets in at the critical field strength of 190 kV/cm, consistent with literature data [14]. The device was fabricated as follows: The active PCM has the shape of a rectangular bar (line) connecting two metallic electrodes. Electrode structures for line-cells were fabricated using optical sapphire substrates. The electrodes consisted of a 10 nm thick adhesion layer of titanium and 120 nm of tungsten in contact with the PCM. To achieve a stable contact resistance to the contacting tips, big platinum pads were deposited on extensions of the tungsten electrodes. A thin films of AIST was deposited using DC-magnetron sputtering from stoichiometric targets in inert argon atmosphere. To avoid contamination, the chamber was evacuated to a background pressure below $10^{-6}$ mbar before sputter depositing the PCM. The native oxide on the tungsten electrodes was removed in situ using an rf-plasma etching process. After the deposition process (in 5.3 mbar Argon atmosphere) the PCM was capped in-situ with a thin 10 nm layer of $ZnS:SiO_2$ to prevent oxidation. The width w and length l (distance between the electrodes) of the device prepared in this way was verified with the help of SEM measurements. The thickness t of the films was calculated based on the calibrated deposition rate.

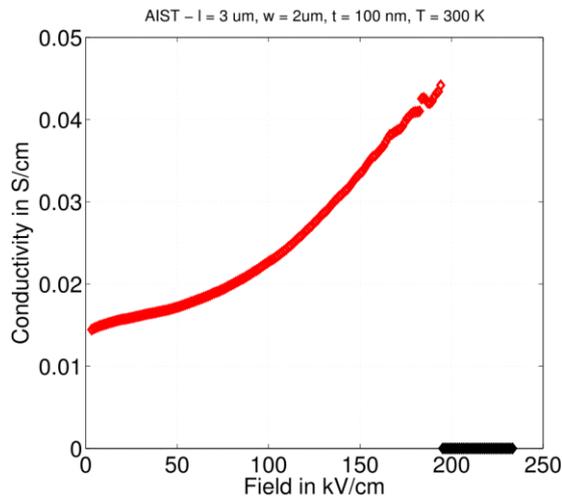

**Fig. S13:** The DC conductivity of amorphous AIST is field dependent, but in the sub-threshold regime depicted here, the conductivity changes only by a factor of three. Data were obtained on line cells with geometry explained in the text.

**References**

[1] H. Y. Hwang, S. Fleischer, N. C. Brandt, B. G. Perkins, M. Liu, K. Fan, A. Sternbach, X. Zhang, R. D. Averitt, and K. A. Nelson, "A review of non-linear terahertz spectroscopy with ultrashort tabletop-laser pulses," *J. Mod. Opt.*, vol. 62, no. 18, pp. 1447–1479, 2015.

[2] P. C. M. Planken, H.-K. Nienhuys, H. J. Bakker, and T. Wenckebach, "Measurement and calculation of the orientation dependence of terahertz pulse detection in ZnTe," *J. Opt. Soc. Am. B*, vol. 18, no. 3, p. 313, 2001.

[3] M. Naftaly, Ed., *Terahertz Metrology*. Artech House, 2015.

[4] I. Wilke and S. Sengupta, "Nonlinear Optical Techniques for Terahertz Pulse Generation and Detection—Optical Rectification and Electrooptic Sampling," in *Terahertz Spectroscopy*, S. Dexheimer, Ed. CRC Press, 2008.

[5] S. Casalbuoni, H. Schlarb, B. Schmidt, B. Steffen, P. Schmüser, and a. Winter, "Numerical studies on the electro-optic sampling of relativistic electron bunches," *Proc. IEEE Part. Accel. Conf.*, vol. 2005, pp. 3070–3072, 2005.

[6] K. Strössner, S. Ves, and M. Cardona, "Refractive index of GaP and its pressure dependence," *Phys. Rev. B*, vol. 32, no. 10, pp. 6614–6619, 1985.

[7] M. Shu, P. Zalden, F. Chen, B. Weems, I. Chatzakis, F. Xiong, R. Jeyasingh, M. Hoffmann, E. Pop, P. Wong, M. Wuttig, and A. Lindenberg, "Ultrafast electric-field-induced processes in GeSbTe phase-change materials," *Appl. Phys. Lett.*, vol. 104, p. 251907, 2014.

[8] F. Kadlec, C. Kadlec, and P. Kužel, "Contrast in terahertz conductivity of phase-change materials," *Solid State Commun.*, vol. 152, no. 10, pp. 852–855, 2012.

[9] P. Zalden, A. von Hoegen, P. Landreman, M. Wuttig, and A. M. Lindenberg, "How supercooled liquid phase-change materials crystallize: Snapshots after femtosecond optical excitation," *Chem. Mater.*, vol. 27, no. 16, p. 5641, 2015.

[10] K. Fan, H. Y. Hwang, M. Liu, A. C. Strikwerda, A. Sternbach, J. Zhang, X. Zhao, X. Zhang, K. a. Nelson, and R. D. Averitt, "Nonlinear terahertz metamaterials via field-enhanced carrier dynamics in GaAs," *Phys. Rev. Lett.*, vol. 110, no. 21, pp. 1–5, 2013.

[11] D. J. Cook and R. M. Hochstrasser, "Intense terahertz pulses by four-wave rectification in air.," *Opt. Lett.*, vol. 25, no. 16, pp. 1210–2, Aug. 2000.

[12] D. Grischkowsky, S. Keiding, M. van Exter, and C. Fattinger, "Far-infrared time-domain spectroscopy with terahertz beams of dielectrics and semiconductors," *J. Opt. Soc. Am. B*, vol. 7, no. 10, p. 2006, 1990.

[13] J. L. M. Oosthoek, D. Krebs, M. Salinga, D. J. Gravesteijn, G. a M. Hurkx, and B. J. Kooi, "The influence of resistance drift on measurements of the activation energy of conduction for phase-change material in random access memory line cells," *J. Appl. Phys.*, vol. 112, no. 8, p. 4506, 2012.

[14] D. Krebs, S. Raoux, C. T. Rettner, G. W. Burr, M. Salinga, and M. Wuttig, "Threshold field of phase change memory materials measured using phase change bridge devices," *Appl. Phys. Lett.*, vol. 95, no. 8, p. 2101, 2009.

[15] K. Lai, W. Kundhikanjana, M. Kelly, and Z. X. Shen, "Modeling and characterization of a cantilever-based near-field scanning microwave impedance microscope," *Rev. Sci. Instrum.*, vol. 79, no. 6, p. 3703, 2008.

[16] W. Kundhikanjana, K. Lai, H. Wang, H. Dai, M. a. Kelly, and Z. X. Shen, "Hierarchy of electronic properties of chemically derived and pristine graphene probed by microwave imaging," *Nano Lett.*, vol. 9, no. 11, pp. 3762–3765, 2009.

[17] P. Jost, "Charge Transport in Phase-Change Materials," PhD thesis, RWTH Aachen University, 2013.

[18] S. G. Bishop, P. C. Taylor, D. L. Mitchell, and L. H. Slack, "Far infrared and microwave conductivity spectrum of semiconducting Tl2Se·As2Te3 glass," *J. Non. Cryst. Solids*, vol. 5, no. 4, pp. 351–357, 1971.

[19] W. K. Njoroge, M. Wuttig, and I. Introduction, "Crystallization kinetics of sputter-deposited amorphous AgInSbTe lms," *J. Appl. Phys.*, vol. 90, no. 8, pp. 3816–3821, 2001.

[20] N. F. Mott and E. A. Davis, *Electronic processes n non-crystalline materials*. Oxford: Clarendon Press, 1979.